\documentclass[12pt]{iopart}
\usepackage{iopams} 
\usepackage{setstack} 
\usepackage[dvips]{graphicx} 
\usepackage{color} 
\usepackage{epsfig} 
\usepackage{epstopdf}
\usepackage{hyperref}
\usepackage{subfig}
\begin{document}

\title[Post-Newtonian expansion of a disc of charged dust]{Post-Newtonian
expansion of a rigidly rotating disc of dust with a constant specific charge}

\author{Stefan Palenta and Reinhard Meinel}

\address{Theoretisch-Physikalisches Institut, Friedrich-Schiller-Universit\"{a}t
Jena, Max-Wien-Platz 1, 07743 Jena, Germany}
\eads{\mailto{stefan.palenta@uni-jena.de}, \mailto{meinel@tpi.uni-jena.de}} 

\begin{abstract}
We present an algorithm for obtaining the post-Newtonian expansion
of the asymptotically flat solution to the Einstein-Maxwell equations
describing a rigidly rotating disc of dust with a constant specific charge.
Explicit analytic expressions are calculated up to the eighth order. The results
are used for a physical discussion of the extreme relativistic limiting cases.
We identify strong evidence for a transition to an extreme Kerr-Newman black
hole. 
\end{abstract}

\submitto{\CQG}

\section{Introduction}

Since the formulation of classical general relativity by Einstein in 1915,
analytic solutions have made a large contribution to the understanding of its
meanings and proving its properties. A crucial role in this process was played
by black hole solutions, as their description requires the full generality of
Einstein's theory. Therefore black hole solutions could be expected to exhibit
inherent effects of general relativity. Indeed, the Schwarzschild solution
containing a single mass parameter $M$ was able to explain the perihelion
precession of Mercury and the bending of light rays around the sun. The Kerr
solution extended the class of black hole solutions by the parameter of angular
momentum $J$, led to the discovery of ergospheres and permitted the rigorous
investigation of the Lense-Thirring effect. Furthermore, adding a charge
parameter $Q$ completes the class of stationary black hole solutions by
delivering the most general Kerr-Newman solution, see \cite{noHair1, noHair2}.

Interestingly, static spherically symmetric ideal fluid configurations have a
minimal radius, given by $\frac{9}{8}$ times the corresponding Schwarzschild
radius and thus fail to feature a continuous connection to black holes
\cite{Buchdahl1959}. The analytic solution of the problem of a rigidly rotating
uncharged disc of dust, in contrast, provides an example for a parametric
transition of equilibrium configurations of ordinary matter to an extreme Kerr
black hole \cite{Neugebauer1995}, see also \cite{RelFig}. Rotating fluid rings
admit such a black hole limit too \cite{Dyson, BHL}.

\begin{figure}
\centering
\includegraphics[width=0.6\linewidth]{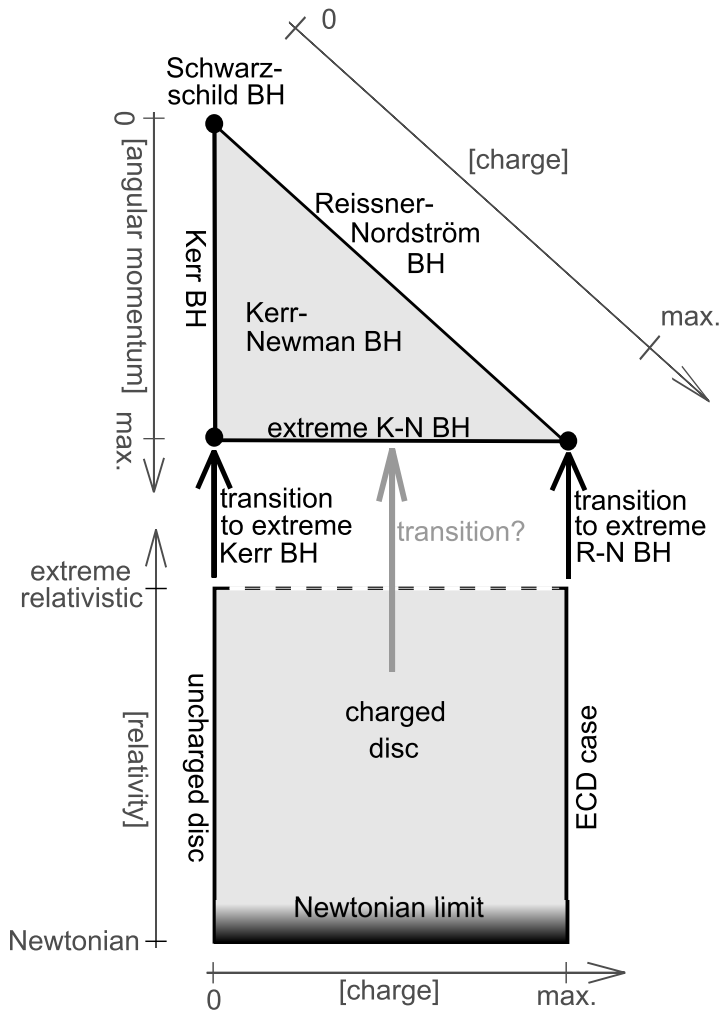}
\caption{Schematic of the parameter spaces of black holes and charged dust discs
linked by two limiting cases.}
\label{paramSpaceExt}
\end{figure}

On the other hand, the so called electrically counterpoised dust (ECD)
configurations, a class of static solutions of dust with a particular constant
specific charge, possess a parametric transition to the extreme
Reissner-Nordstr\"{o}m black hole, see \cite{Meinel2011}. Inspired by these
opposite limiting cases, we conjecture that the dust disc with a constant
specific charge varying from zero to the ECD value has a parametric transition
to the extreme Kerr-Newman black hole. The parameter spaces of black holes and
charged dust discs and their connections are illustrated in
\fref{paramSpaceExt}. In order to investigate this assumption and to support the
search for the analytic solution of the charged disc, a post-Newtonian expansion
is performed.
We note that in the uncharged case, the analytic disc solution
\cite{Neugebauer1995} was indeed preceded by the partly numerical but highly
accurate Bardeen-Wagoner expansion \cite{BardWag}, which could later be
recovered from the analytic solution \cite{Petroff2001}. The charged disc is
expected to show novel effects due to the interplay between electromagnetic and
gravitational fields.
\section{Formulation of the disc problem}
\subsection{Model of matter}

To facilitate analytic calculations, we employed the following simplifying
assumptions:

Firstly, we use the model of dust, i.e.\ pressureless matter described by the
dust part of the energy momentum tensor in \eref{Tab}. Moreover, we assume a
rigid rotation with a constant angular velocity $\Omega$. The dust we consider
here contains the constant specific charge $\epsilon\in[-1,1]$ so that the
charge density $\rho_{\rm el}$ and the baryonic mass density $\mu$ are connected
via 
\begin{equation}
\rho_{\rm el}=\epsilon \mu
\end{equation} 
(we use Gauss units and $c=1,\, G=1$).

Furthermore, we assume axial symmetry and stationarity described by the Killing
vectors $\boldsymbol\eta$ and $\boldsymbol\xi$, as well as reflectional symmetry
w.r.t. an ``equatorial plane''.

\subsection{Basic field equations}

We start with Einstein's field equations
\begin{equation} 
 R_{ab} - \frac{1}{2}Rg_{ab} = 8\pi T_{ab}
\end{equation}
and the covariant Maxwell equations
\begin{equation}  
 F_{[ab;c]}=0,\qquad F^{ab}_{\quad;b}=4\pi \jmath^a  \label{maxwell_gl},
\end{equation}
wherein the energy momentum tensor $T_{ab}$ is simply the sum of an
electromagnetic part and a dust part:
\begin{equation}  \label{Tab}
 T_{ab} = T^{{(\rm em})}_{ab} + T^{({\rm dust})}_{ab} = 
\frac{1}{4\pi}\left(F_{ac}F_b^{\;c}-\frac{1}{4}g_{ab}F_{cd}F^{cd}\right) + \mu
u_au_b.
\end{equation}
For the purely convective four-current density holds $\jmath^a=\rho_{\rm
el}u^a=\epsilon \mu u^a$. The metric can globally be written in terms of
Weyl-Lewis-Papapetrou coordinates as:
\begin{equation}  
\mathrm{d}s^2=f^{-1}\left[h\left(\mathrm{d}\varrho^2+\mathrm{d}
\zeta^2\right)+\varrho^2\mathrm{d}\varphi^2\right]-f\left(\mathrm{d}t+a\mathrm{d
}\varphi\right)^2 .
\end{equation}
Moreover, a four-potential of the form $A_a=(0,0,A_\varphi,A_t)$ can be
introduced by $F_{ab} = A_{b;a}-A_{a;b} = A_{b,a}-A_{a,b}$, which automatically
fulfils the homogeneous Maxwell equations.
The coordinates correspond to the Killing vectors via
$\partial_t=\boldsymbol\xi$ and $\partial_\varphi=\boldsymbol\eta$, so that all
the five free functions  $f$, $h$, $a$, $A_t$ and $A_\varphi$ depend only on the
coordinates $\varrho$ and $\zeta$. In addition, $\mu$ is made up from a surface
mass density by 
\begin{equation}
\mu=\sigma_{(0)}(\varrho)\delta(\zeta).
\end{equation}
Note that two other definitions of surface mass density will later be used:
$\sigma=\sigma_{(0)}h/f$, directly linked with the boundary conditions, and the
coordinate independent proper surface mass density $\sigma_{\rm
p}=\sigma\sqrt{f/h}$.

The coupled nonlinear system of differential equations that emerge can be
reformulated as a boundary value problem for the equations valid in the
(electro-)vacuum outside the disc. 

\subsection{Ernst equations}

In the absence of matter, the Einstein-Maxwell equations can be reduced to the
Ernst equations. They are expressed in two additional potentials $\beta$ and
$b$. Using the abbreviations $A_t=-\alpha$ and $A_\phi=A$, these potentials are
defined by
\begin{equation} \label{D1}
\beta_{,\varrho} =  \frac{f}{\varrho}\left(a \alpha_{,\zeta}+ A_{,\zeta}\right),
\qquad
\beta_{,\zeta}   = -\frac{f}{\varrho}\left(a \alpha_{,\varrho}+
A_{,\varrho}\right), 
\end{equation}
and
\begin{equation} \label{D2}
b_{,\varrho} = -\frac{f^2}{\varrho}a_{,\zeta} + 2\left( \beta \alpha_{,\varrho}
- \alpha \beta_{,\varrho} 					
\right),\qquad
b_{,\zeta}   =\,\frac{f^2}{\varrho}a_{,\varrho}+ 2\left(\beta \alpha_{,\zeta} -
\alpha\beta_{,\zeta} \right).
\end{equation}
Thereby the complex Ernst potentials  $\mathcal{E}$ and $\Phi$ can be
introduced:
\begin{equation}
\Phi = \alpha + {\rm i}\beta,\qquad  \mathcal{E} = \left(f-
\bar{\Phi}\Phi\right) + {\rm i}b.
\end{equation}
In these terms the Ernst equations read \cite{Ernst}
\begin{eqnarray}
\left(\Re\mathcal{E}+\bar{\Phi}\Phi\right)\Delta\mathcal{E} =\;
\left(\nabla\mathcal{E}+2\bar{\Phi}\nabla\Phi\right)\cdot\nabla\mathcal{E}, \\
\left(\Re\mathcal{E}+\bar{\Phi}\Phi\right)\Delta\Phi =\;
\left(\nabla\mathcal{E}+2\bar{\Phi}\nabla\Phi\right)\cdot\nabla\Phi,
\end{eqnarray}
where the behaviour of $\nabla$ and $\Delta$ is analogous to that in Euclidean
3-space, with cylindrical coordinates $(\varrho,\zeta,\varphi)$. The metric
function $h$ is eliminated during this process of transforming the field
equations to the Ernst equations and can be determined by integration
afterwards. For the purpose of the post-Newtonian expansion, it is helpful to
express the Ernst equations in terms of the real functions  $f$, $\alpha$,
$\beta$ and $b$, which leads to
\begin{eqnarray}
f\Delta f=\,f\Delta\left(\alpha^2\right)+f\Delta\left(\beta^2\right)+\nabla
f\left(\nabla f-2\alpha\nabla\alpha-2\beta\nabla\beta\right)
\nonumber\\\qquad\qquad
-\nabla\beta\left(\nabla\beta+2\alpha\nabla\beta-2\beta\nabla\alpha\right),
\label{E1}\\
f\Delta b=\,2\nabla f\nabla
b+4\alpha\beta\left(\left(\nabla\alpha\right)^2-\left(\nabla\beta\right)^2\right
) -4\left(\alpha^2-\beta^2\right)\nabla\alpha\nabla\beta \nonumber\\\qquad\qquad
+2\nabla f\left(\alpha\nabla\beta-\beta\nabla\alpha\right)-2\nabla
b\left(\alpha\nabla\alpha+\beta\nabla\beta\right), \label{E2}\\
f\Delta \alpha=\nabla f\nabla \alpha-\nabla\beta\left(\nabla
b+2\alpha\nabla\beta-2\beta\nabla\alpha\right), \label{E3}\\
f\Delta \beta=\nabla f\nabla\beta+\nabla\alpha\left(\nabla
b+2\alpha\nabla\beta-2\beta\nabla\alpha\right). \label{E4}
\end{eqnarray}
\subsection{Boundary conditions on the disc}

To describe the influence of the matter, we have to return to the full
Einstein-Maxwell equations. We evaluate them in the corotating frame with
$\varphi'=\varphi-\Omega t$, where the metric retains its form:
\begin{equation} \label{Metrik'}
\mathrm{d}s^2={f'}^{-1}\left[h'\left(\mathrm{d}\varrho^2+\mathrm{d}
\zeta^2\right)+\varrho^2\mathrm{d}{\varphi'}^2\right]-f'\left(\mathrm{d}
t+a'\mathrm{d}\varphi'\right)^2 
\end{equation}
with the dashed functions related to the original ones by
\begin{equation} \label{trafo-bezieh}
 \fl f' =  f\left[\left(1+\Omega a\right)^2 -
\frac{\Omega^2\varrho^2}{f^2}\right],\quad
 \left(1+\Omega a\right)f = \left(1-\Omega a'\right)f' \quad {\rm and}\quad
 \frac{h}{f} = \frac{h'}{f'}.
\end{equation}
The Einstein-Maxwell equations in the corotating frame lead to the system
\begin{eqnarray}
0 =- \nabla\cdot\left[\frac{f'}{\varrho^2}\left(a'\nabla \alpha '+\nabla
A'\right) \right], \label{Well_I_rot2}\\
4\pi\sigma\epsilon\,{f'}^{-\frac{1}{2}}\delta\left(\zeta\right) =
\nabla\cdot\left[\frac{a'f'}{\varrho^2}\left(a'\nabla \alpha '+\nabla
A'\right)-\frac{1}{f'}\nabla \alpha '\right], \label{Well_II_rot2}\\
0 = \nabla\cdot\left[\frac{{f'}^2}{\varrho^2}\nabla a' +
4\frac{f'}{\varrho^2}\alpha '\left(a'\nabla \alpha '+\nabla A'\right) \right],
\label{FGL-II_rot2}\\
\fl 8\pi\sigma {f'}^2\delta\left(\zeta\right)  = f'\Delta f' - \left(\nabla
f'\right)^2 + \frac{{f'}^4}{\varrho^2}\left(\nabla a'\right)^2 
- 2f'\left[\left(\nabla \alpha '\right)^2 +
\frac{{f'}^2}{\varrho^2}\left(a'\nabla \alpha '+\nabla A'\right)^2 \right],
\label{FGL-I_rot2}\\
\fl \left(\ln h'\right)_{,\varrho} = \frac{1}{2}\varrho\left[{\left(\ln
f'\right)_{,\varrho}}^2-{\left(\ln f'\right)_{,\zeta}}^2 -
\frac{{f'}^2}{\varrho^2}\left({{a'}_{,\varrho}}^2-{{a'}_{,\zeta}}^2\right)\right
] + 2\left[\frac{f'}{\varrho}\left({{A'}_{,\varrho}}^2-{{A'}_{,\zeta}}^2\right)
\right. \nonumber\\
\left. -\frac{\varrho^2-{a'}^2{f'}^2}{f'\varrho}\left({{\alpha
'}_{,\varrho}}^2-{{\alpha
'}_{,\zeta}}^2\right)+2\frac{a'f'}{\varrho}\left(A'_{,\varrho}\alpha
'_{,\varrho}-A'_{,\zeta}\alpha '_{,\zeta}\right)\right],
\label{FGL-h_rho_rot2}\\
\fl \left(\ln h'\right)_{,\zeta} =\varrho\left[4\left(\ln
f'\right)_{,\varrho}\left(\ln f'\right)_{,\zeta} -
\frac{{f'}^2}{\varrho^2}a'_{,\varrho}a'_{,\zeta}\right] \nonumber\\
 + 4\left[ \frac{f'}{\varrho}\left(A'_{,\varrho}+a'\alpha
'_{,\varrho}\right)\left(A'_{,\zeta}+a'\alpha '_{,\zeta}\right) -
\frac{\varrho}{f'}\alpha '_{,\varrho}\alpha '_{,\zeta} \right].
\label{FGL-h_zeta_rot2}
\end{eqnarray}

Integration over a small flat cylinder around a mass element of the disc (very
analogous to the well known treatment of surface charge densities in
electrostatics) delivers matching conditions between the areas above and beneath
the disc. These matching conditions are transformed to the following four
boundary conditions by the use of reflectional symmetry:
\begin{equation}
\beta'=0,\qquad b'=0,\qquad
\alpha '_{,\zeta} = -\epsilon\left({f'}^{\frac{1}{2}}\right)_{,\zeta},\qquad
\left({f'}^{\frac{1}{2}}\right)_{,\varrho} = -\epsilon \alpha '_{,\varrho}.
\end{equation}
Back in the nonrotating frame they take the form:
\begin{eqnarray}
\left(\frac{\Omega\varrho^2}{(1+\Omega a)f^2}-a\right)\alpha_{,\zeta} =
A_{,\zeta}, \label{R1}\\
\Omega\varrho^2\left(\frac{\Omega\varrho^2}{(1+\Omega a)f^2}\right)_{,\zeta}=
a_{,\zeta}, \label{R2}\\
\epsilon\left(\sqrt{\left(1+\Omega
a\right)^2f-\Omega^2\varrho^2f^{-1}}\right)_{,\zeta} = \left(\Omega
A-\alpha\right)_{,\zeta},  \label{R3}\\
\left(\sqrt{\left(1+\Omega
a\right)^2f-\Omega^2\varrho^2f^{-1}}\right)_{,\varrho} = \epsilon\left(\Omega A
-\alpha\right)_{,\varrho}.  \label{R4}
\end{eqnarray}
The treatment in the nonrotating frame is justified by the simple boundary
conditions at spatial infinity, which are given by $g_{ab}\to\eta_{ab}$ and
$A_a\to 0$ in that system. 

Overall we have the 4 differential equations \eref{E1}--\eref{E4} and the 4
boundary conditions \eref{R1}--\eref{R4} for the 6 expansion functions $f$,
$\alpha$, $\beta$, $b$, $A$ and $a$. However, due to the definitions \eref{D1}
and \eref{D2}, only 4 of these functions are independent and therefore the disc
problem is well posed.

\section{Ansatz for the expansion}
\subsection{Parameter space of the charged disc}

\begin{figure} 
\centering 
\includegraphics[width=0.5\linewidth, trim=0 2mm 0 2mm]{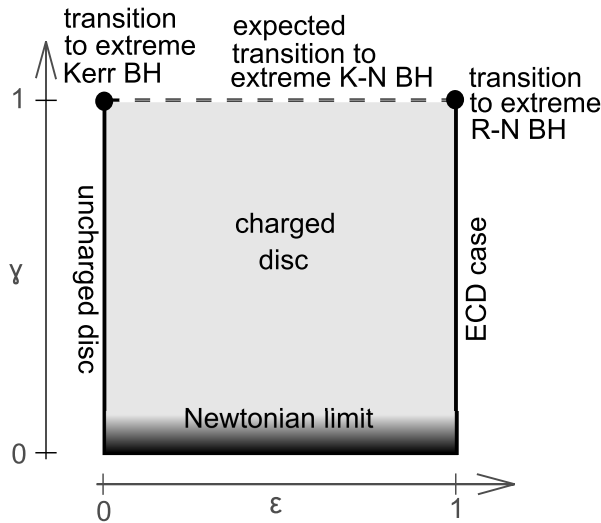}
\caption{Parameter space of the charged disc.}
\label{paramSpace}
\end{figure}

We introduce the relativity parameter $\gamma$, which was already successfully
used for the post-Newtonian expansion of the uncharged disc \cite{BardWag}:
\begin{equation} \label{DefG}
\gamma=1-\sqrt{f_{\rm c}}\qquad {\rm with}\quad f_{\rm c}=f(\varrho=0,\zeta=0).
\end{equation}
The parameter $\gamma$ is closely related to the redshift $Z_{\rm c}$ of a
photon emitted at the center of the disc and measured at infinity:
\begin{equation} 
\gamma=Z_{\rm c}(1+Z_{\rm c})^{-1}.
\end{equation}
Therefore, it is not surprising that the transition to a black hole happens at
$\gamma\to 1$ for the uncharged and the ECD case. The parameter space of the
charged disc as shown in \fref{paramSpaceExt} (w.l.o.g. restricted to positive
charges) is now strongly conjectured to be the area $\gamma\in[0,1]$,
$\epsilon\in[0,1]$ shown in \fref{paramSpace}. 
As a third parameter, the disc radius $\varrho_0$ is added. It naturally comes
into play by introducing elliptic coordinates through 
\begin{equation*}
\zeta=\varrho_0\xi\eta \qquad{\rm and}\qquad
\varrho=\varrho_0\sqrt{(1-\eta^2)(1+\xi^2)}
\end{equation*}
and carries the whole information about the scale of the disc. By the usage of
elliptic coordinates and by multiplying powers of $\varrho_0$ to unit carrying
quantities, the whole disc problem can be made dimensionless. Those normalized
quantities, labeled with a star, are for example
\begin{equation}
a^*=a/\varrho_0,\qquad A^*=A/\varrho_0, \qquad \Omega^*=\varrho_0\Omega \quad
{\rm and}\qquad \sigma^*=\varrho_0\sigma.
\end{equation}

\subsection{Formulation of power series}

Inspired by the Newtonian disc of charged dust (in the Newtonian limit we have
$\gamma=-U_{\rm N}(\varrho=0,\zeta=0)=\Omega^{*2}/(1-\epsilon^2)$ with the
Newtonian potential $U_{\rm N}$), the squared angular velocity is expanded as a
power series in $\gamma$:
\begin{equation}
\Omega^{*2}=(1-\epsilon^2)\gamma+\mathcal{O}\left(\gamma^{2}\right).
\end{equation}
The angular velocity $\Omega^*$ is therefore described by an odd series in
$\gamma^{\frac{1}{2}}$:
\begin{equation}\label{OmegEntw}
\Omega^*=\sqrt{\gamma}\sum_{n=1}^\infty\Omega^*_n\gamma^{n-1}. 
\end{equation}

All the previously discussed functions are now either symmetric or antisymmetric
with respect to a change of the sense of rotation $\Omega^*\to-\Omega^*$. As
such, they can be expressed either in even or odd power series of
$\gamma^{\frac{1}{2}}$. The lowest occuring power is determined by the Newtonian
limit. On the whole we get
\begin{eqnarray} 
\eqalign{
f=1+\gamma\sum_{n=1}^\infty f_n\gamma^{n-1}, \qquad
\alpha=\gamma\sum_{n=1}^\infty \alpha_n\gamma^{n-1}, \qquad
a=\gamma^{\frac{3}{2}}\sum_{n=1}^\infty a_n\gamma^{n-1}, \\
A=\gamma^{\frac{3}{2}}\sum_{n=1}^\infty A_n\gamma^{n-1},\qquad
\beta=\gamma^{\frac{3}{2}}\sum_{n=1}^\infty \beta_n\gamma^{n-1}, \qquad
 b=\gamma^{\frac{3}{2}}\sum_{n=1}^\infty b_n\gamma^{n-1}.
 }
\end{eqnarray}
The remaining coefficients are still functions of $\xi$ and $\eta$. These
expansions are inserted in the 12 equations \eref{E1}--\eref{E4},
\eref{R1}--\eref{R4} plus \eref{D1} and \eref{D2}. These 12 equations, which we
call expansion equations, are now evaluated by equating coefficients of
powers of $\gamma$ in order to obtain analytic expressions.

\section{Algorithmic solution}
\subsection{Structural examination}

The expansion equations in $k$-th order read:
\numparts
\begin{eqnarray}
(1+\xi^2)\beta_{k,\xi} = -A_{k,\eta}/\varrho_0+ F_{\rm D1k} \label{D1k}\\
(1-\eta^2)\beta_{k,\eta} = +A_{k,\xi}/\varrho_0+ F_{\rm D2k}\label{D2k} \\
(1+\xi^2)b_{k,\xi} = +a_{k,\eta}/\varrho_0+ F_{\rm D3k}  \label{D3k}\\
(1-\eta^2)b_{k,\eta} = -a_{k,\xi}/\varrho_0+ F_{\rm D4k}  \label{D4k}
\end{eqnarray}
\endnumparts
\numparts
\begin{eqnarray}
\sqrt{1-\epsilon^2}(1-\eta^2)\alpha_{k,\xi} = +A_{k,\xi}/\varrho_0+ F_{\rm B1k}
\label{R1k}\\
2\sqrt{1-\epsilon^2}(1-\eta^2)f_{k,\xi}= -a_{k,\xi}/\varrho_0+ F_{\rm B2k}
\label{R2k}\\
\epsilon f_{k,\xi} = -2\alpha_{k,\xi}+ F_{\rm B3k} \label{R3k}\\
4\sqrt{1-\epsilon^2}\eta\varrho_0\Omega_k+f_{k,\eta} =
-2\epsilon\alpha_{k,\eta}+ F_{\rm B4k}  \label{R4k}
\end{eqnarray}
\endnumparts
\numparts
\begin{eqnarray}
\Delta f_k= F_{\rm E1k} \label{E1k}\\
\Delta b_k= F_{\rm E2k} \label{E2k}\\
\Delta \alpha_k= F_{\rm E3k}\label{E3k}\\
\Delta \beta_k= F_{\rm E4k} \label{E4k}
\end{eqnarray}
\endnumparts
For clarity, only coefficient functions of $k$-th order have been written down,
and terms of lower orders in each equation are subsumed in the Symbol $F$. These
functions $F$ all vanish for $k=1$ so that the first order equations reduce to
coupled Laplace equations with boundary conditions. The gravitational and the
electric potential of the Newtonian disc are reproduced through $f_1=2U_{\rm N}$
and $\alpha_1=-U_{\rm el}$ (see \eref{NewtPot}). For $k>1$, none of the
functions $F$ vanish and we get coupled Poisson equations. The requirement for
asymptotical flatness implies that all the 6 coefficient functions $f_k$,
$\alpha_k$, $\beta_k$, $b_k$, $A_k$ and $a_k$ have to vanish at spatial infinity.

\subsection{Solving the boundary value problem}

A very helpful observation comes from the fact that all expansion functions are
polynomials in $\eta$. Therefore the functions $F$ (besides a factor of
$(\xi^2+\eta^2)^{-1}$ in the $F_{\rm Eik}$) are also polynomials in $\eta$ and
the Poisson inhomogeneity can thus be fragmented to Legendre polynomials. Then
the Poisson equation with an inhomogeneity consisting of a single Legendre
polynomial,
\begin{equation}
\Delta\psi=\frac{1}{\xi^2+\eta^2}
\left\lbrace\left[(1+\xi^2)\psi_{,\xi}\right]_{,\xi}+\left[(1-\eta^2)\psi_{,\eta
}\right]_{,\eta}\right\rbrace
=\frac{I(\xi)P_n(\eta)}{\xi^2+\eta^2},
\end{equation}
is transformed by the ansatz $\psi=A(\xi)P_n(\eta)$ into the ordinary
differential equation
\begin{equation}
\left[(1+\xi^2)A_{,\xi}\right]_{,\xi}-n(n+1)A=I(\xi).
\end{equation}
Two independent (real) solutions to the corresponding homogeneous equation are 
\begin{equation}
A_1={\rm i}^n P_n({\rm i}\xi),\qquad A_2={\rm i}^{1-n} Q_n({\rm i}\xi)
\end{equation}
where $Q_n$ are the Legendre functions of the second kind. Now the
inhomogeneous solution can be composed as 
\begin{equation}
\fl A(\xi)=-A_1(\xi)\int_{c_1}^\xi \frac{A_2(x)I(x)}{(1+\xi^2)W} {\rm
d}x+A_2(\xi)\int_{c_2}^\xi \frac{A_1(x)I(x)}{(1+\xi^2)W}{\rm d}x,\quad W=A'_2
A_1-A'_1 A_2.
\end{equation}
With adequately chosen constants $c_1$ and $c_2$, the boundary conditions at
$\xi\to\infty$ and $\xi\to 0$ can be fulfilled respectively.

\subsection{Completion of the expansion equations and algorithmic solution}

In addition to the 12 expansion equations \eref{D1k}--\eref{E4k}, two parameter
relations are necessary to determine an integration constant and the scalar
component $\Omega_k$ for each order. First, from \eref{DefG}, we immediately
obtain at $\varrho=0$, $\zeta=0$
\begin{equation}
f_{1{\rm c}}=-2,\quad f_{2{\rm c}}=1 \quad{\rm and}\quad f_{k{\rm c}}=0 \quad
{\rm for}\;\; k>2. \label{FC}
\end{equation}
Secondly, the following relation to the surface mass density can be derived
analogously to the derivation of the boundary conditions ($F_{...}$ still
collects lower order terms):
\begin{equation} 
\frac{1}{\eta}\left[f_{k,\xi}+ F_{\rm REG1k}\right] =
8\pi\varrho_0\sigma_k+ F_{\rm REG2k} \label{REGk}
\end{equation}
In order to have a finite value of $\sigma_k$ at the rim of the disc, $\eta=0$,
the expression $\left[f_{k,\xi}+ F_{\rm REG1k}\right]$ has to contain a global
factor $\eta$. The integration constants of the other unknown functions vanish due to
reflectional symmetry.

The system of expansion equations can be dealt with by setting up the linear
combination $O_k=\frac{1}{2}\epsilon f_k+\alpha_k$. This provides a quantity
with a given Poisson inhomogeneity and a given boundary condition:
\begin{equation}
\Delta O_k=\frac{\epsilon}{2} F_{\rm E1k}+F_{\rm E3k}, \qquad
O_{k,\xi}=\frac{1}{2} F_{\rm B3k}.
\end{equation}
Once $O_k$ is determined according to 4.2, $f_k$ can be determined by \eref{R4k} and $\alpha_k$ is
immediately obtained. Hereupon $\beta_k$ and $b_k$ can be computed successively
out of Poisson boundary value problems and $A_k$ and $a_k$
eventually by integration.

\section{Results}

\begin{figure} 
\centering
 \includegraphics[width=0.6\linewidth, trim=0 0mm 0 0mm]{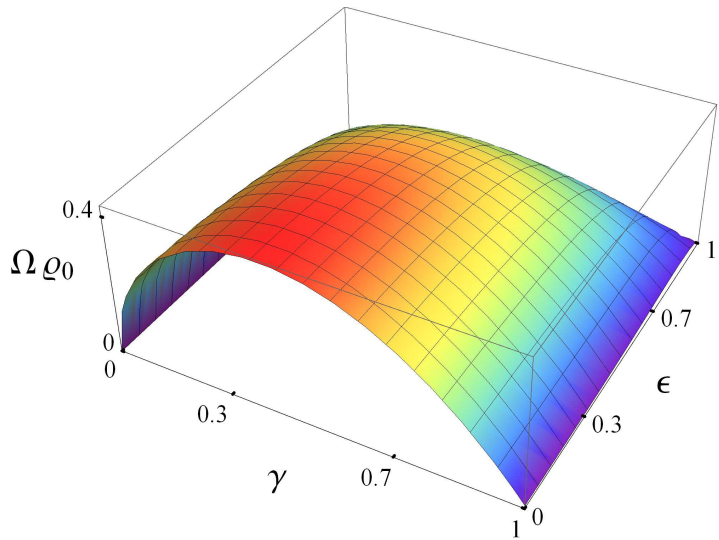} 
 \caption{Plot of $\Omega^*$, the angular velocity normalized by the disc
radius.}
 \label{OmegaPlot}
\end{figure}

With the methods described above and making use of Mathematica by Wolfram
Research, coefficient functions of the 6 expansion functions could be calculated
up to the eighth order. The first important observation is that these
coefficient functions are not only polynomial in $\eta$ but also in $\xi$ and
${\rm arccot}\,\xi$. The first 2 orders are listed in the appendix, higher
orders are available as data files. Noteworthy are the global prefactors in the
expansion functions, which could be found in all of their coefficients:
\begin{eqnarray}
\eqalign{
\Omega^*_k\propto\sqrt{1-\epsilon ^2},\quad \alpha_k\propto\epsilon,\quad
\beta_k\propto\epsilon\sqrt{1-\epsilon^2}\eta,\quad
b_k\propto\sqrt{1-\epsilon^2}\eta,\\
 A^*_k\propto\epsilon\sqrt{1-\epsilon^2}(1-\eta^2) \quad{\rm and}\quad
a^*_k\propto\sqrt{1-\epsilon^2}(1-\eta^2).
 }
\end{eqnarray}
Within the expansion, the angular velocity $\Omega^*$ is also calculated. It is
depicted in \fref{OmegaPlot} up to the eighth order.
Here the parameter space (\fref{paramSpace}) is shown in the $\gamma$-$\epsilon$
plane.
$\Omega^*$ vanishes at the transition to the static configurations for
$\epsilon\to 1$ and in the extreme relativistic limit ($\gamma\to 1$). The
latter
is caused by the vanishing disc radius $\varrho_0$. This shows that  $\varrho_0$
is no longer a well chosen parameter for $\gamma\to 1$.

\begin{figure}
\centering
\includegraphics[width=0.6\linewidth, trim=0 4mm 0 4mm]{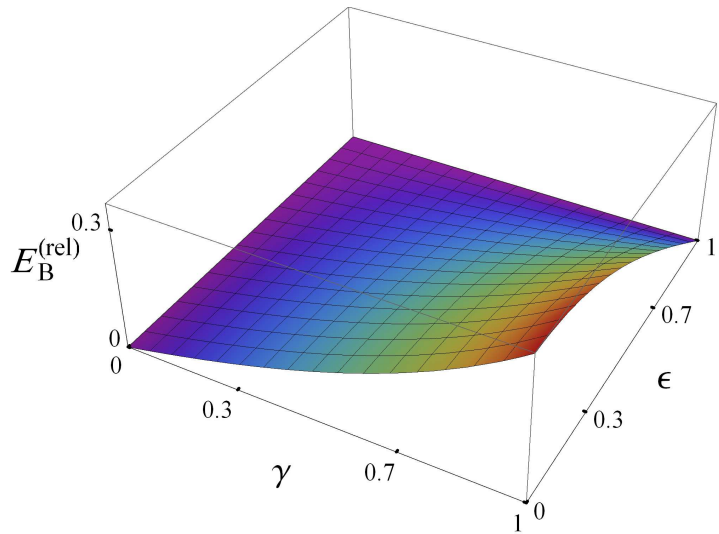}
\caption{Relative binding energy.}
\label{EBrelPlot}
\end{figure}

The further evaluation of the results is done by discussing global quantities of
the disc. The gravitational mass $M$, the charge $Q$ and the angular momentum
$J$ can be conveniently deduced from the far field behaviour of the expansion
functions:
\begin{eqnarray}
f =\; 1 - \frac{2M^*}{\xi}+  \mathcal{O}\left(\xi^{-2}\right),  \\
\alpha =\; \frac{Q^*}{\xi}+  \mathcal{O}\left(\xi^{-2}\right), \\
a =\; \frac{2J^*\left(1-\eta^2\right)}{\xi} +  \mathcal{O}\left(\xi^{-2}\right)
.
\end{eqnarray}
One compelling feature of the disc with constant specific charge, is that even
the baryonic mass $M_0=\epsilon^{-1}Q$ can be derived from the asymptotical
behaviour. Based on this, the relative binding energy $E_{\rm B}^{\rm
(rel)}=(M_0-M)M_0^{-1}$ can be directly calculated (even in the limit
$\epsilon\to 0$). It is depicted in \fref{EBrelPlot} up to the seventh
significant order. The first three orders read
\begin{eqnarray}
\fl E_{\rm B}^{\rm (rel)}=\frac{1}{5} \left(1-\epsilon
^2\right)\gamma+\frac{2}{175} \left(2 \epsilon ^4-9 \epsilon
^2+7\right)\gamma^2 +
\frac{\epsilon ^2-1}{3024000 \pi
   ^2}\left[179200 \left(3 \epsilon ^4-52 \epsilon ^2+64\right)
\right.\nonumber\\
   \left. - \pi ^2 \left(58618 \epsilon ^4-1042821 \epsilon ^2+1275648\right)
\right]\gamma^3 
   +\mathcal{O}\left(\gamma^4\right),
\end{eqnarray}
all seven orders are given in the appendix.
At this point, the accuracy of the expansion can be tested by comparison with
the known analytic solution of the uncharged disc. The potentially most critical
point is $\epsilon\to 0,\;\gamma\to 1$, where the transition to the extreme Kerr
black hole occurs. Here the numerically evaluated expansion gives $E_{\rm
B}^{\rm (rel)}(\epsilon=0,\gamma=1)=0.3614$, which differs from the analytic
value $0.373283588\dots$ \cite{RelFig} by about $3\%$. In the uncharged limit
($\epsilon\to 0$), the expressions for the remaining nontrivial functions $f$
and $a$ agree with an expansion of the analytic solution of the uncharged disc
\cite{Petroff2001}. In addition, the results for the global quantities for
$\epsilon\to 0$ comply with those of the uncharged disc given in \cite{RelFig}.
A non-trivial test for all $\epsilon$ consists in checking the exact relation
\cite{Proceed}
\begin{equation}
M=2\Omega J+(1-\gamma+\epsilon \alpha_{\rm c})M_0, 
\qquad\alpha_{\rm c}=\alpha(\varrho=0,\zeta=0).
\end{equation}
This relation is satisfied by our results for all eight available orders in
$\gamma$.

\begin{figure}[htb] 
  \centering
  \begin{minipage}{0.4\linewidth}
  \subfloat[]{\includegraphics[width=\linewidth, trim= 0 3mm 0
3mm]{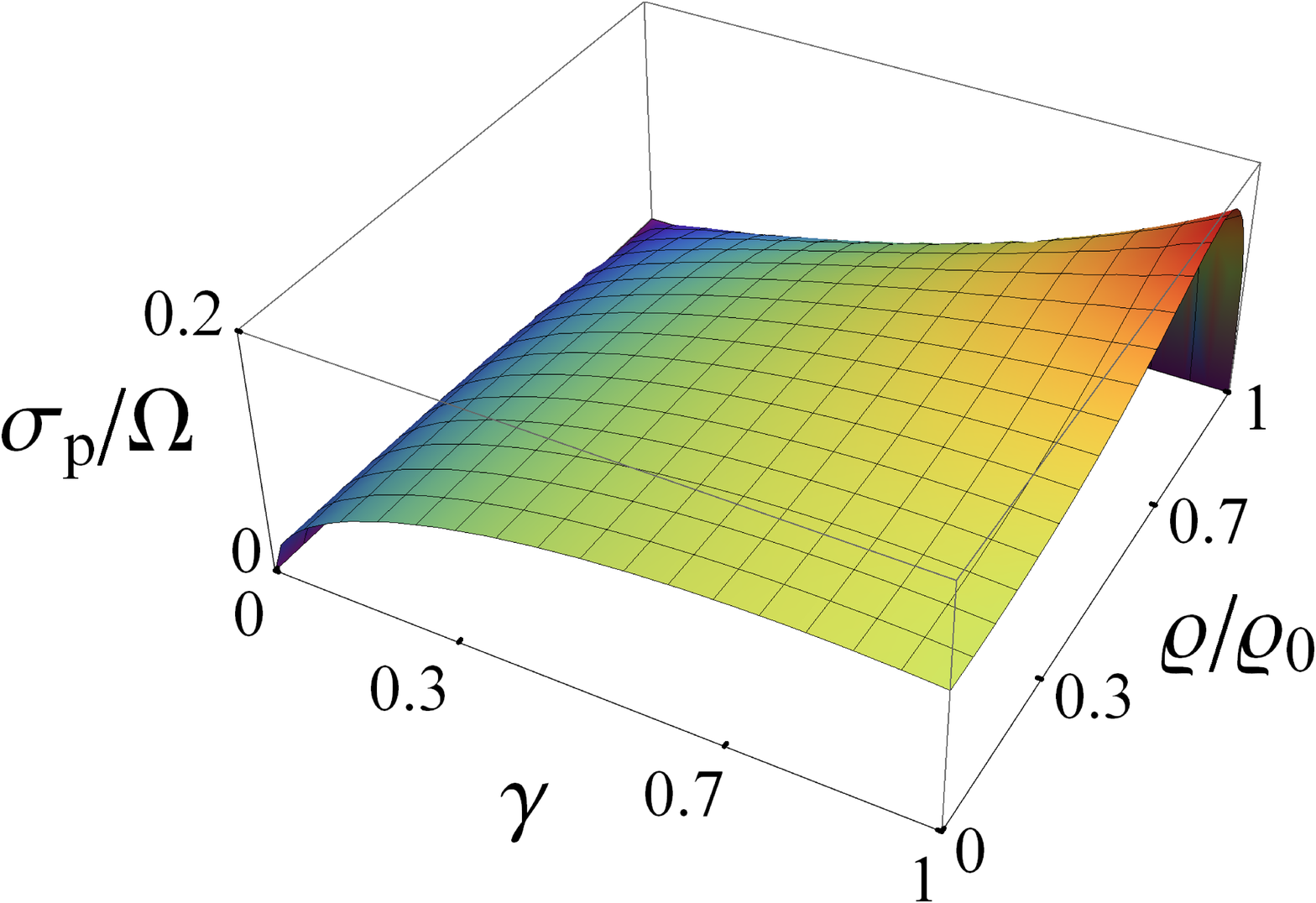}}
  \end{minipage}
  \hspace{0.05\linewidth}
  \begin{minipage}{0.4\linewidth} 
  \subfloat[]{\includegraphics[width=\linewidth, trim= 0 3mm 0
3mm]{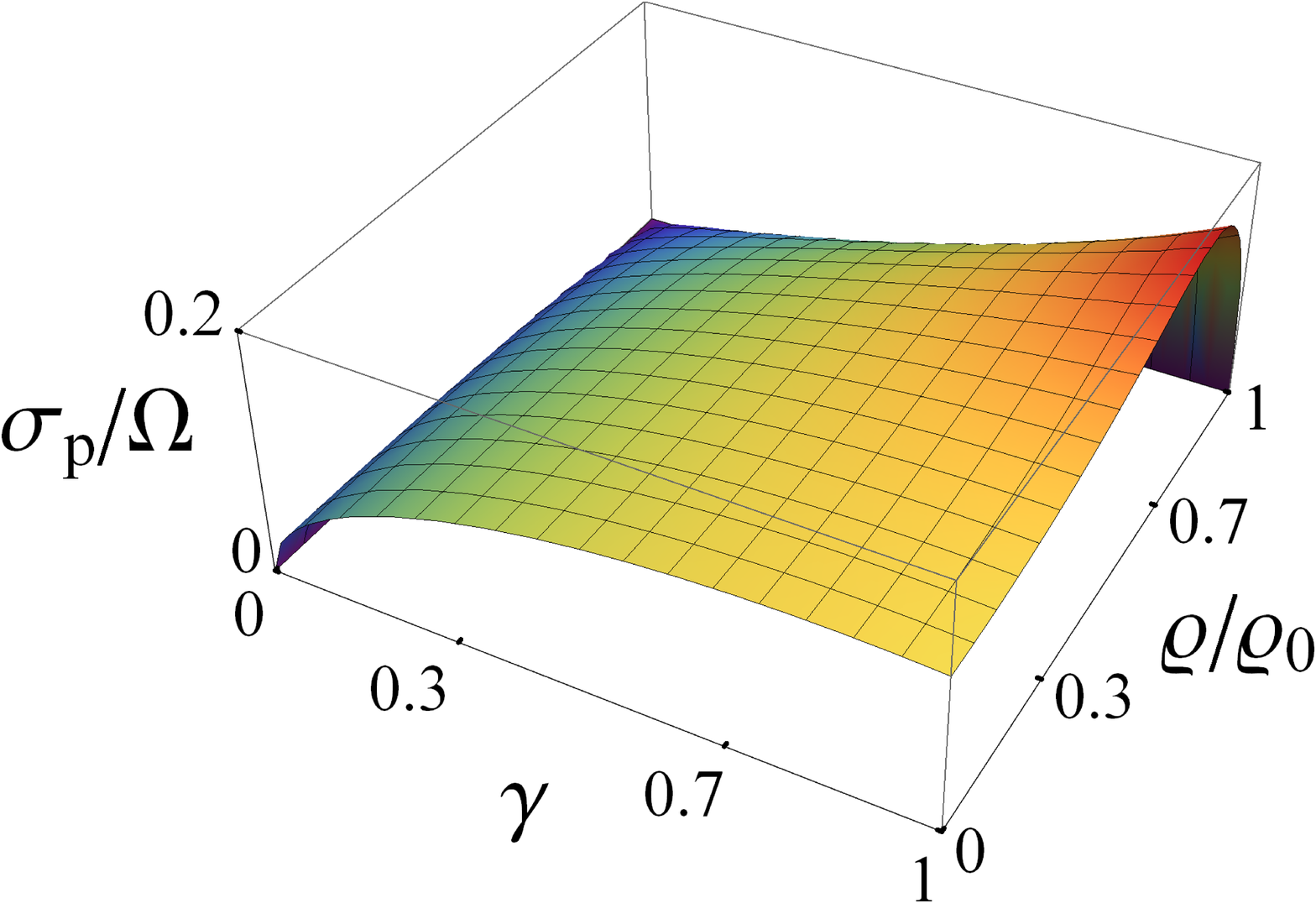}}
  \end{minipage}
  \begin{minipage}{0.4\linewidth}  
  \subfloat[]{\includegraphics[width=\linewidth, trim= 0 3mm 0
3mm]{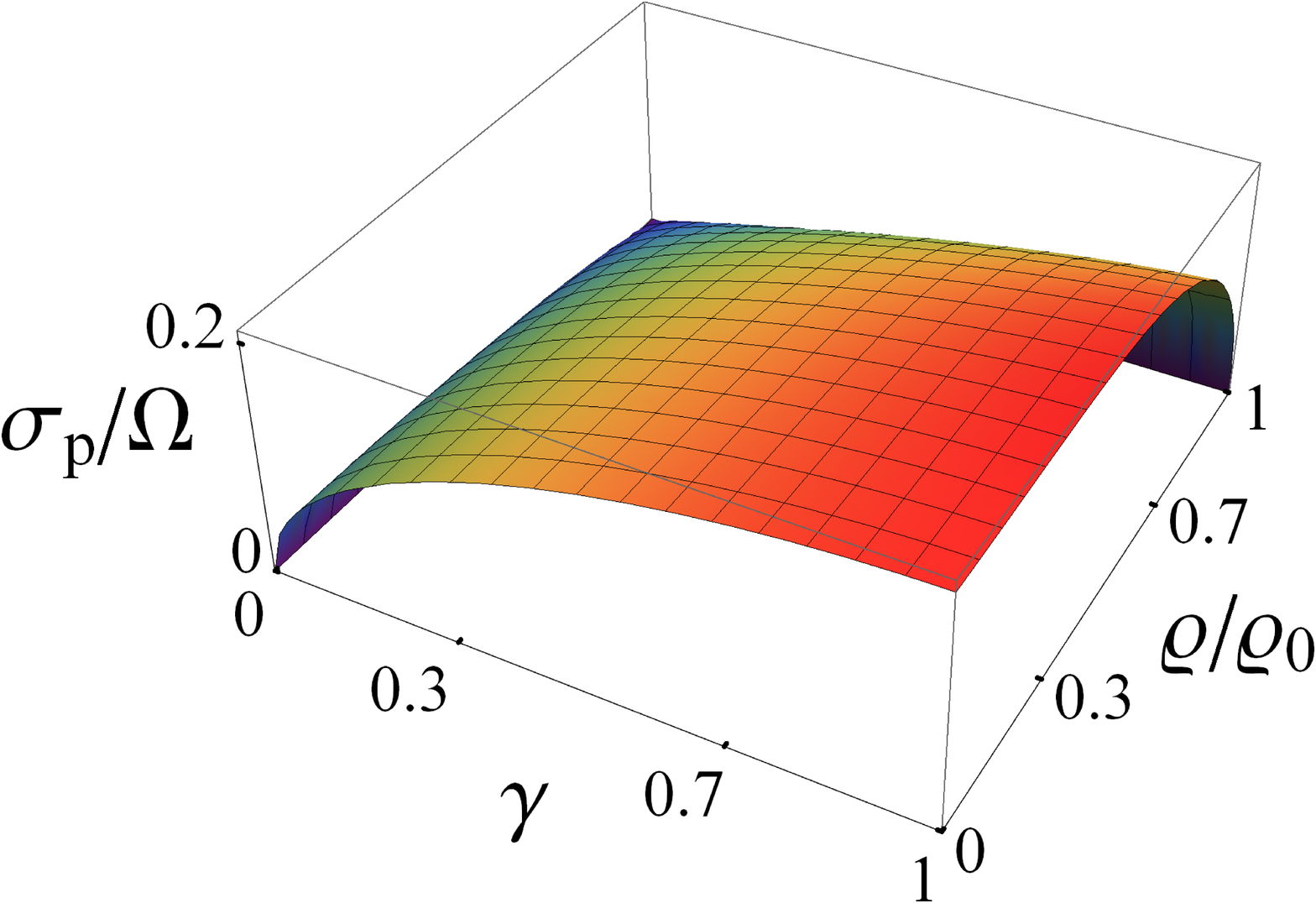}}
  \end{minipage}
  \hspace{0.05\linewidth}
  \begin{minipage}{0.4\linewidth}
  \subfloat[]{\includegraphics[width=\linewidth, trim= 0 3mm 0
3mm]{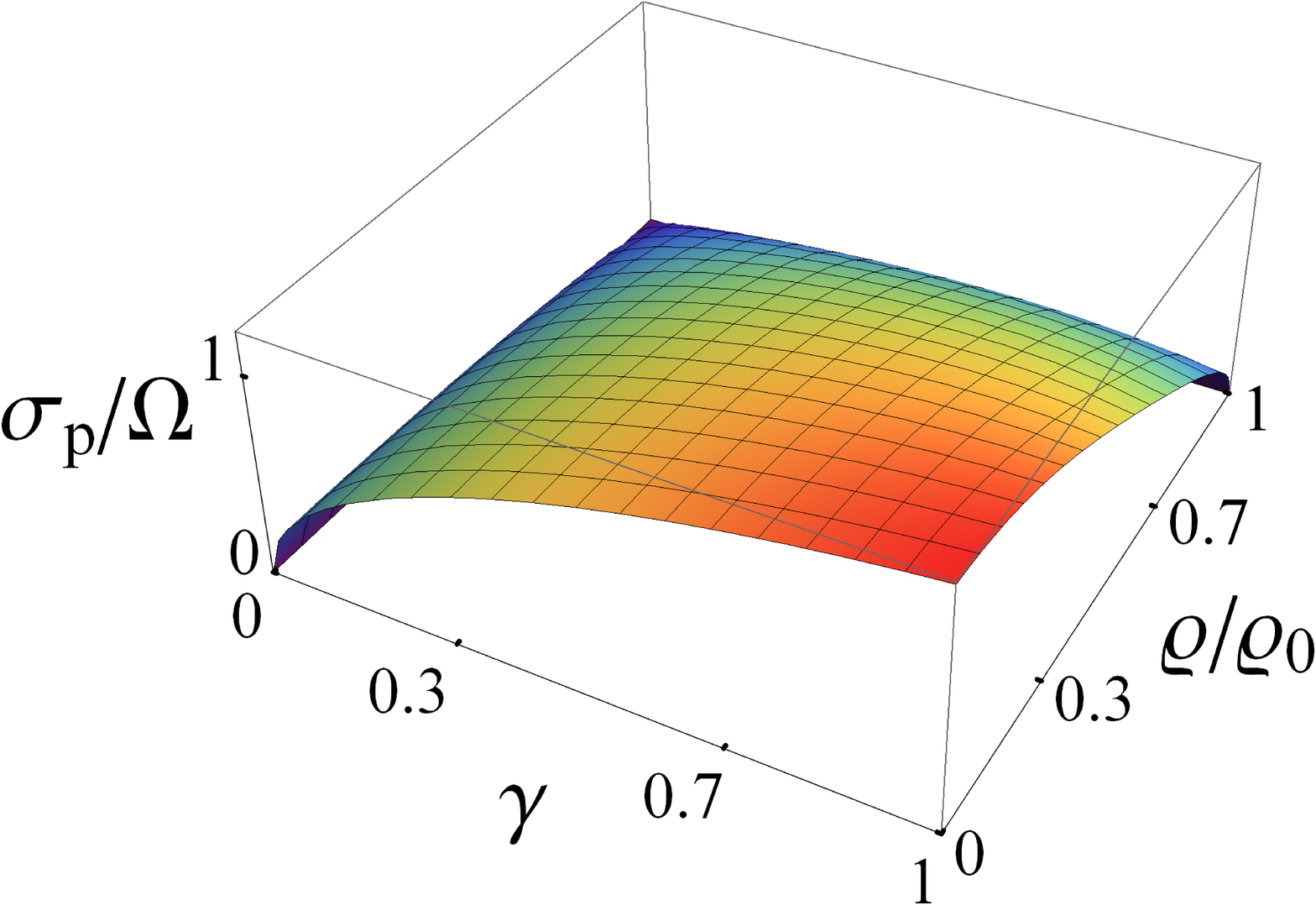}}
  \end{minipage}
  \caption{Surface mass density normalized by angular velocity, $\sigma_{\rm
p}/\Omega$, for specific charges of (a) $\epsilon=0$, (b) $\epsilon=0.3$, (c)
$\epsilon=0.7$, (d) $\epsilon=0.99$.}
  \label{SurfMassPlot}
\end{figure}

\Fref{SurfMassPlot} shows the coordinate independent surface mass density
$\sigma_{\rm p}$ normalized by the angular velocity $\Omega$. In contrast to the
normalization by $\varrho_0$ in \fref{OmegaPlot}, this gives a plot regular at
$\gamma\to 1$. The already known shift of the maximum of $\sigma_{\rm p}/\Omega$
out of the disc centre appears to weaken with growing charge. For $\epsilon\to
1$ the angular velocity $\Omega$ vanishes and is hence no more an appropriate
scaling parameter.

\begin{figure}
  \centering
  \includegraphics[width=0.6\linewidth, trim= 0 1mm 0 1mm]{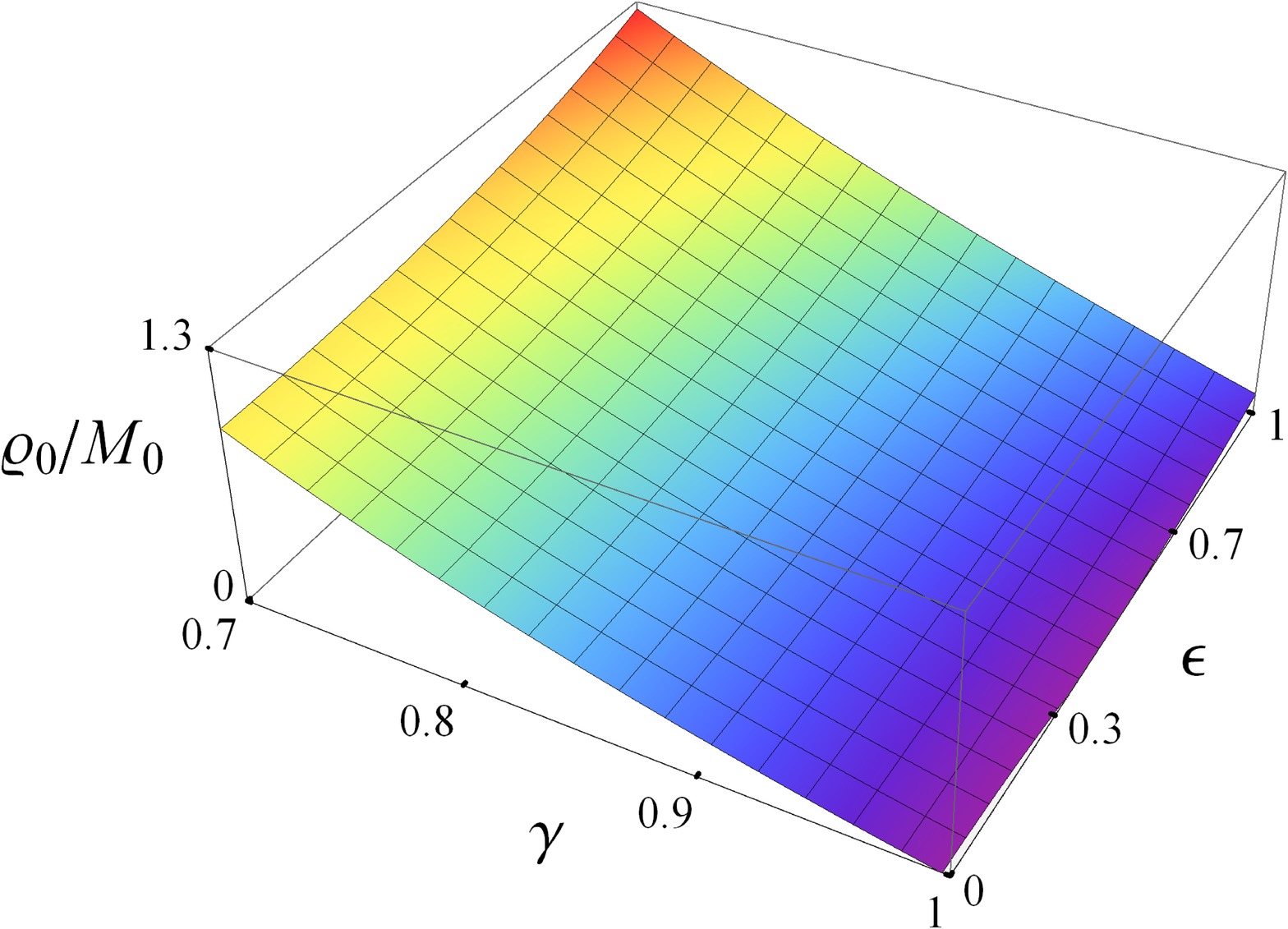}
  \caption{Disc radius $\varrho_0$ normalized by the baryonic mass $M_0$.}
  \label{rho0M0Plot}
\end{figure}

Another interesting quantity is the disc radius $\varrho_0$, normalized by the
baryonic mass $M_0$ which is depicted in \fref{rho0M0Plot}. The small values at
$\gamma\to 1$ and the weak dependence on $\epsilon$ indicate a transition to a
black hole as in the limiting cases $\epsilon=0$ and $\epsilon=1$. Note that one
has to distinguish the external perspective (finite $\varrho/M$, $\zeta/M$) from
the internal perspective (finite $\varrho/\varrho_0$, $\zeta/\varrho_0$) in the
limit $\gamma\to 1$, see the discussion in \cite{Proceed}.

\section{Conclusions}

\begin{figure}
  \centering
  \includegraphics[width=0.6\linewidth, trim= 0 1mm 0 1mm]{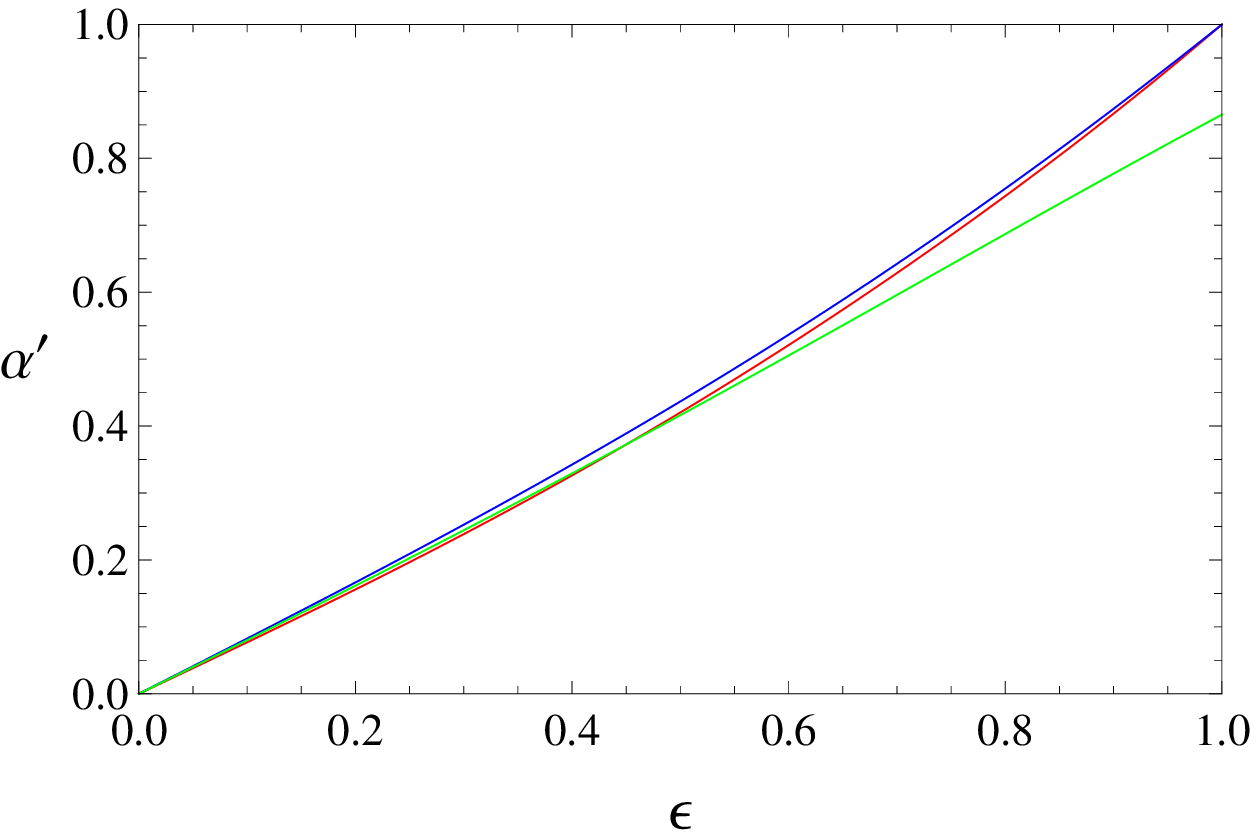}
  \caption{$\alpha '$ calculated via \eref{alphaEBrel} (red) as well as taken
from the disc values at $\varrho^*=0$ (blue) and $\varrho^*=1$ (green) in the
limit $\gamma\to 1$ (expansion order $n=8$).}
  \label{LimAlfs}
\end{figure}

The vanishing of the quantity $\varrho_0/M_0$ for $\gamma\to 1$ can be
interpreted as the contraction of a disc with given baryonic mass to the origin
of the $(\varrho/M,\zeta/M)$-coordinate system in the extreme relativistic
limit. This strongly indicates the occurrence of a singularity in that limiting
case. To check whether this singularity is indeed related to an extreme
Kerr-Newman black hole, three other, more specific tests are carried out. For
the extreme Kerr-Newman metric, the following relation holds on the horizon
$\mathcal{H}$ (located at $\varrho/M=0=\zeta/M$):
\begin{equation} \label{alphaEBrel}
\alpha '(\mathcal{H})=\left[2\frac{M}{Q}-\frac{Q}{M}\right]^{-1}.
\end{equation}
The corresponding $\epsilon$-dependence of $\alpha '(\varrho/M=0,\,\zeta/M=0)$
computed out of $E_{\rm B}^{\rm (rel)}$ via $M/Q=\epsilon^{-1}(1-E_{\rm B}^{\rm
(rel)})$ is depicted in \fref{LimAlfs} as the red line. If we transform to
normalized coordinates by $\varrho\to\varrho^*,\;\zeta\to\zeta^*$ (keep in mind
that we expect $\varrho_0\to 0$ for $\gamma\to 1$), the extreme relativistic
limit yields the limiting spacetime from the internal perspective. The relation
$\alpha '(\varrho^*)=const$ on the whole disc is strongly conjectured to be
necessary and sufficient for obtaining the extreme Kerr-Newman metric as the
external solution (i.e.\ for $(\varrho^2+\zeta^2)/M^2>0$), provided
$\varrho_0\to 0$ as $\gamma\to 1$. The curves of $\alpha '(\varrho^*=0)$ and
$\alpha '(\varrho^*=1)$ are depicted in \fref{LimAlfs} in blue and green.
The similarity of the three curves shows that $\alpha'$ is indeed in good
approximation independent of $\varrho^*$ and that its value meets the one given
by \eref{alphaEBrel}. 

\begin{figure}
  \centering
  \subfloat[]{\includegraphics[width=0.48\linewidth]{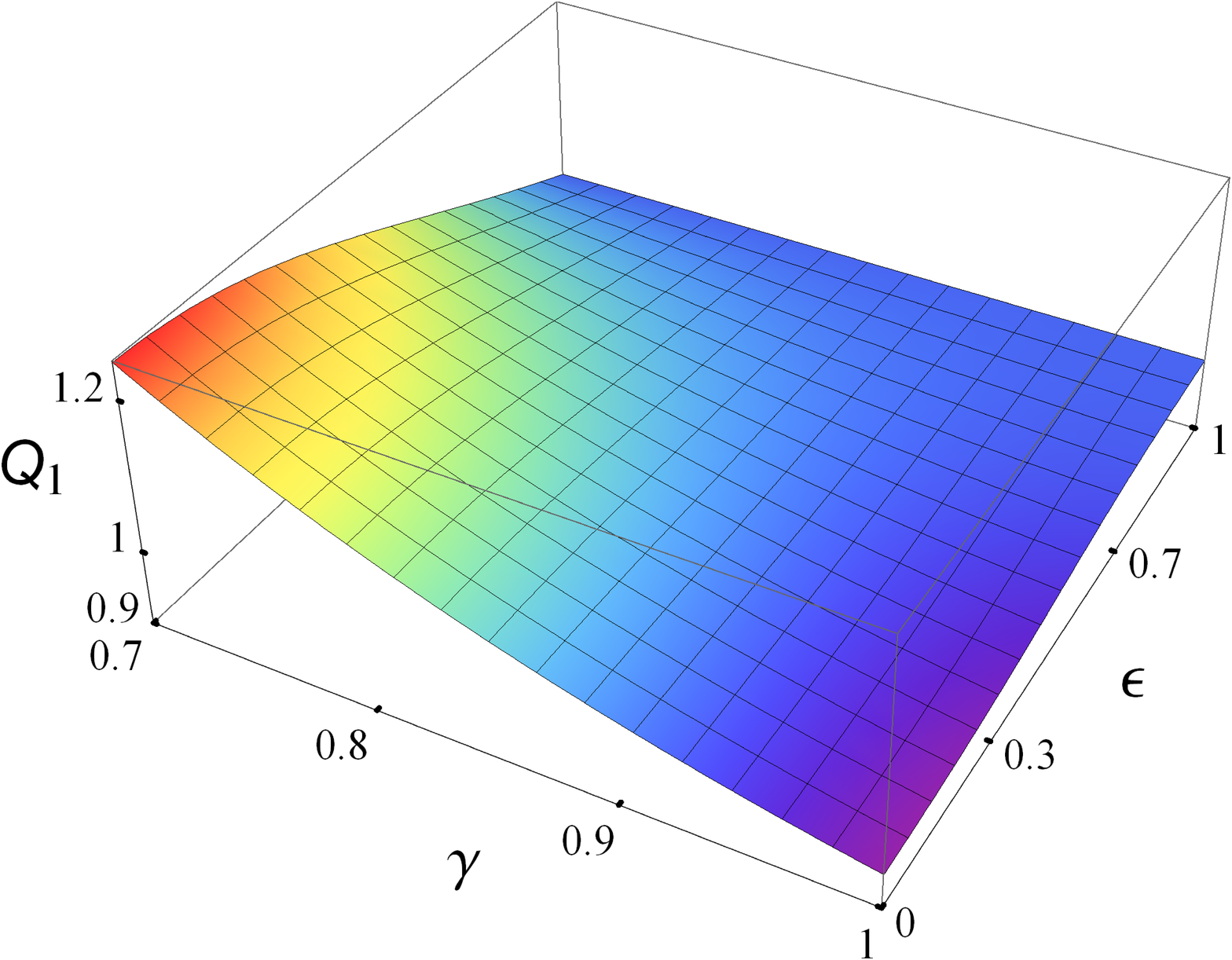}}
  \hspace{0.03\linewidth}
 
\subfloat[]{\includegraphics[width=0.48\linewidth,height=0.38\linewidth]{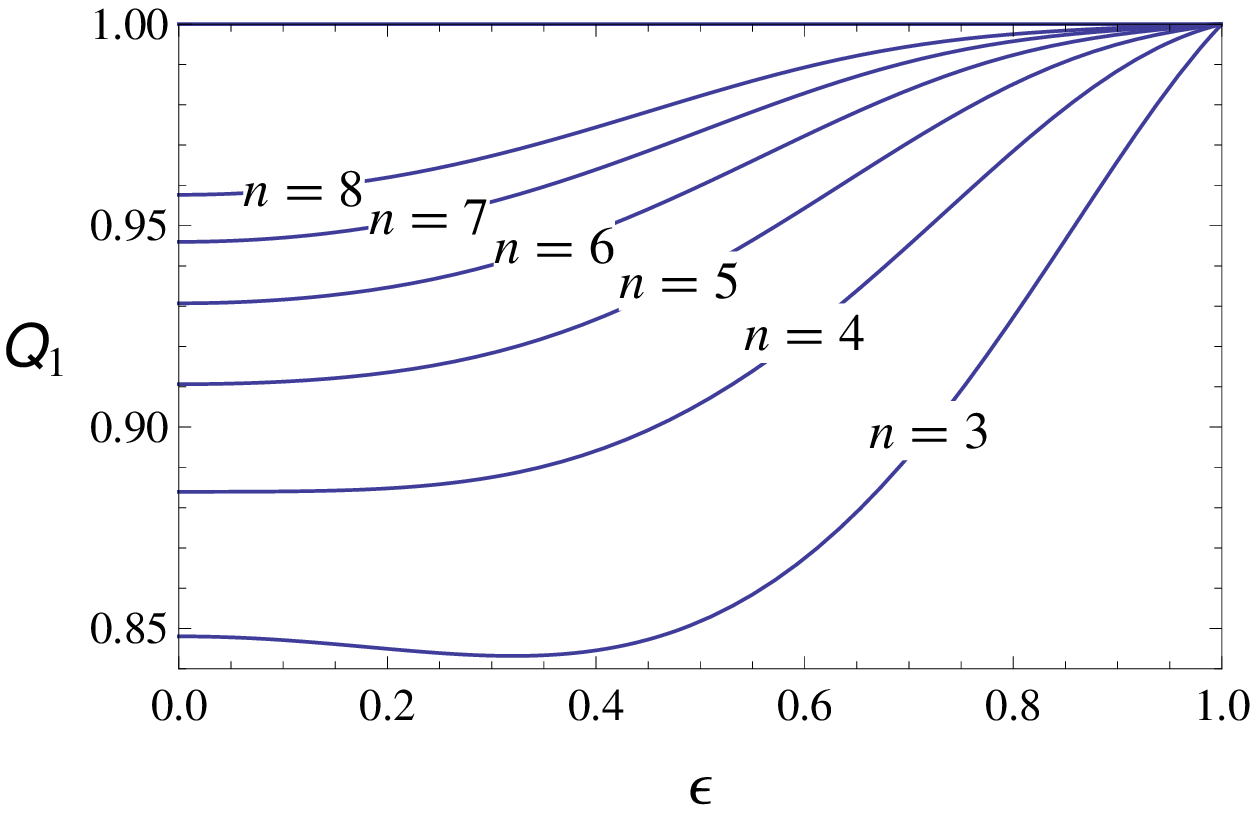}
}
  \caption{Plots of $Q_1$ for the charged disc with $n=8$ (a) and in the limit
$\gamma\to 1$ for different expansion orders $n$ (b).}
  \label{Q1}
\end{figure}

\begin{figure}
  \centering
  \includegraphics[width=0.6\linewidth]{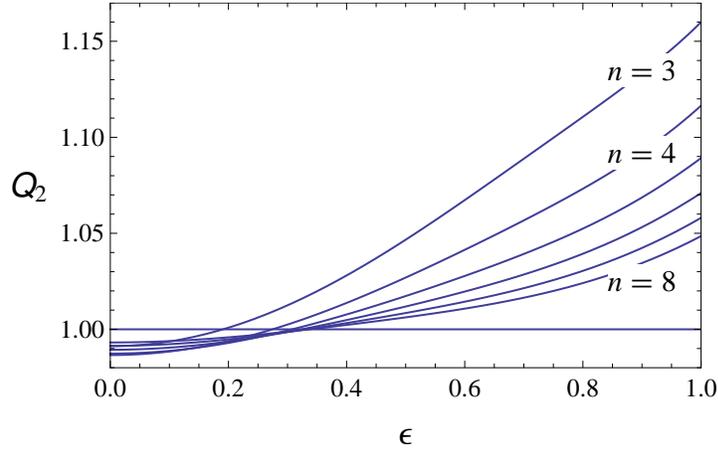}
  \caption{Plots of $Q_2$ in the limit $\gamma\to 1$ for different expansion
orders $n$.}
  \label{Q2}
\end{figure}

Furthermore the extreme Kerr-Newman black hole obeys the 2 parameter relations
\begin{equation}
Q_1:=\frac{Q^2}{M^2}+\frac{J^2}{M^4}=1\qquad {\rm and} \qquad
Q_2:=\frac{J/M^2}{\Omega M(1+J^2/M^4)}=1
\end{equation}
with $\Omega$ meaning the angular velocity of the horizon. The first of these
quotients, calculated out of the disc quantities, is shown in \fref{Q1}. 
For the generic disc $Q_1$ is greater than 1 as expected, whereas in the extreme
relativistic limit $Q_1$ converges to unity for growing expansion order $n$. The
last statement also holds for the second quotient $Q_2$ shown in \fref{Q2}. 
Therefore we conclude that we have found strong evidence for the transition of
the charged disc to an extreme Kerr-Newman black hole in the limit $\gamma\to
1$.

\ack
This research was supported by the Deutsche Forschungsgemeinschaft (DFG) through
the Graduiertenkolleg 1532 ``Quantum and Gravitational Fields''. The authors
would like to thank Martin Breithaupt, Yu-Chun Liu and Christopher J. Pynn for
valuable discussions.
\appendix
\setcounter{section}{0}

\section{Coefficient functions in first order}

In the appendix we use the abbreviation $X:={\rm arccot}\,\xi$. Global factors
are exposed at the cost of showing the fragmentation to Legendre polynomials.
\begin{equation} 
\Omega_1^*=\sqrt{1-\epsilon^2}
\end{equation}
\begin{equation}\label{NewtPot}
\fl f_1=-\frac{2}{\epsilon}\alpha_1=-\frac{8}{3\pi}X
-\frac{4}{3\pi} \left[\frac{3\xi^2+1}{2}X-\frac{3}{2}\xi\right](3\eta^2-1)
\end{equation}
\begin{equation}
\fl \beta_1=-\frac{\epsilon}{4}b_1 =\frac{ \epsilon\sqrt{1-\epsilon ^2} }{3 \pi
} \eta  \left[9 \xi ^2-(15 \xi ^2+4)\eta ^2+3 \xi  \left( (5 \xi^2+3)\eta ^2-3
\xi ^2-1\right) X \right]
\end{equation}
\begin{eqnarray}
\fl a^*_1=\frac{4}{\epsilon}A^*_1=-\frac{\sqrt{1-\epsilon ^2}}{\pi }(\eta ^2-1)
\left[\xi -13 \eta ^2 \xi+(3-15 \eta ^2) \xi ^3 \right.\nonumber\\ \left. +(\xi
^2+1) \left(3 \eta ^2 (5 \xi ^2+1)-3 \xi ^2+1\right) X \right].
\end{eqnarray}

\section{Coefficient functions in second order}

\begin{equation} 
\Omega_2^*=\sqrt{1-\epsilon^2}\frac{2\epsilon^2-9}{12}
\end{equation}

\begin{eqnarray}
\fl f_2=\,\frac{1}{36 \pi ^2}\left[
\left(-72 \pi +15 \pi  \epsilon ^2\right) \xi +\left(72+36 \epsilon ^2\right)
\xi^2+\left(-108 \pi +153 \pi  \epsilon ^2\right) \xi ^3 \right.\nonumber\\
\left.
+\left[-72 \pi +15 \pi  \epsilon ^2+\left(144+72 \epsilon^2\right) \xi
+\left(108 \pi -66 \pi  \epsilon ^2\right) \xi^2  \right.\right.\nonumber\\
\left.\left.
+\left(-144-72 \epsilon ^2\right) \xi ^3+\left(108 \pi -153 \pi  \epsilon
^2\right) \xi ^4\right] X+\left[72+36 \epsilon^2 \right.\right.\nonumber\\
\left.\left.
+\left(-144-72 \epsilon ^2\right) \xi ^2+\left(72+36 \epsilon ^2\right) \xi
^4\right] X^2+\eta ^2 \left\{\left(612 \pi -606\pi  \epsilon ^2\right) \xi
\right.\right.\nonumber\\ \left.\left.
   +\left(-432-216 \epsilon ^2\right) \xi ^2+\left(1080 \pi -1530 \pi  \epsilon
^2\right) \xi^3+ \left[-108 \pi +66 \pi  \epsilon ^2
\right.\right.\right.\nonumber\\ \left.\left.\left. 
   +\left(-288-144 \epsilon ^2\right) \xi +\left(-972 \pi +1116 \pi  \epsilon
^2\right) \xi^2+\left(864+432 \epsilon ^2\right) \xi ^3
\right.\right.\right.\nonumber\\ \left.\left.\left. 
   +\left(-1080 \pi +1530 \pi  \epsilon ^2\right) \xi ^4\right] X+\left[144+72
\epsilon
   ^2+\left(288+144 \epsilon ^2\right) \xi ^2  \right.\right.\right.\nonumber\\
\left.\left.\left. 
   +\left(-432-216 \epsilon ^2\right) \xi ^4\right] X^2\right\}+\eta ^4
\left\{\left(-660   \pi +935 \pi  \epsilon ^2\right) \xi
\right.\right.\nonumber\\ \left.\left.
   +\left(648+324 \epsilon ^2\right) \xi ^2+\left(-1260 \pi +1785 \pi  \epsilon
^2\right)   \xi ^3+\left[108 \pi -153 \pi  \epsilon ^2
\right.\right.\right.\nonumber\\ \left.\left.\left.
   +\left(-432-216 \epsilon ^2\right) \xi +\left(1080 \pi -1530 \pi  \epsilon
^2\right) \xi ^2+\left(-1296-648 \epsilon ^2\right) \xi ^3
\right.\right.\right.\nonumber\\ \left.\left.\left.
   +\left(1260 \pi -1785 \pi  \epsilon ^2\right) \xi ^4\right]
X+\left[72+36\epsilon ^2+\left(432+216 \epsilon ^2\right) \xi
^2\right.\right.\right.\nonumber\\ \left.\left.\left.
   +\left(648+324 \epsilon ^2\right) \xi ^4\right] X^2\right\}
\right],
\end{eqnarray}
\begin{eqnarray}
\fl \alpha_2=\,\frac{\epsilon }{72 \pi ^2}\left[
\left(81 \pi -24 \pi  \epsilon ^2\right) \xi -72 \xi ^2+\left(27 \pi -72 \pi 
\epsilon ^2\right) \xi ^3+\left[81 \pi -24 \pi\epsilon ^2-144 \xi
\right.\right.\nonumber\\ \left.\left.
+\left(-90 \pi +48 \pi  \epsilon ^2\right) \xi ^2+144 \xi ^3+\left(-27 \pi +72
\pi  \epsilon ^2\right) \xi^4\right] X+\left[-72+144 \xi ^2
\right.\right.\nonumber\\ \left.\left.
-72 \xi ^4\right] X^2+\eta ^4 \left\{\left(165 \pi -440 \pi  \epsilon ^2\right)
\xi -648 \xi^2+\left(315 \pi -840 \pi  \epsilon ^2\right) \xi ^3
\right.\right.\nonumber\\ \left.\left. +\left[-27 \pi +72 \pi  \epsilon ^2+432
\xi +\left(-270 \pi +720 \pi \epsilon ^2\right) \xi ^2+1296 \xi ^3+\left(-315
\pi \right.\right.\right.\right.\nonumber\\ \left.\left.\left.\left.
+840 \pi  \epsilon ^2\right) \xi ^4\right] X+\left[-72-432 \xi ^2-648
\xi^4\right] X^2\right\}+\eta ^2 \left\{\left(-342 \pi
\right.\right.\right.\nonumber\\ \left.\left.\left.
+336 \pi  \epsilon ^2\right) \xi +432 \xi ^2+\left(-270 \pi +720 \pi \epsilon
^2\right) \xi ^3+\left[90 \pi -48 \pi  \epsilon ^2+288 \xi
\right.\right.\right.\nonumber\\ \left.\left.\left.
 +\left(432 \pi -576 \pi  \epsilon ^2\right) \xi ^2-864 \xi^3+\left(270 \pi -720
\pi  \epsilon ^2\right) \xi ^4\right] X\right.\right.\nonumber\\ \left.\left.
 +\left(-144-288 \xi ^2+432 \xi ^4\right] X^2\right\}
\right],
\end{eqnarray}
\begin{eqnarray}
\fl \beta_2=\,\frac{\epsilon  \sqrt{1-\epsilon ^2} }{1440 \pi ^2}\eta \left[
\left(9225 \pi -5520 \pi  \epsilon ^2\right) \xi ^2-12960 \xi ^3+\left(-7425 \pi
-7200 \pi  \epsilon ^2\right) \xi
   ^4\right.\nonumber\\ \left.
   +\left[\left(-3735 \pi +1200 \pi  \epsilon ^2\right) \xi -8640 \xi
^2+\left(-6750 \pi +7920 \pi  \epsilon ^2\right) \xi^3+25920 \xi ^4
\right.\right.\nonumber\\ \left.\left.
   +\left(7425 \pi +7200 \pi  \epsilon ^2\right) \xi ^5\right] X+\left[4320 \xi
+8640 \xi ^3-12960 \xi ^5\right]X^2\right.\nonumber\\ \left.
   +\eta ^4 \left\{-2112 \pi -2048 \pi  \epsilon ^2-17280 \xi +\left(-24255 \pi
-23520 \pi  \epsilon ^2\right) \xi ^2-64800 \xi^3\right.\right.\nonumber\\
\left.\left.
   +\left(-31185 \pi -30240 \pi  \epsilon ^2\right) \xi ^4+\left[5760+\left(7425
\pi +7200 \pi  \epsilon ^2\right) \xi +77760\xi ^2
\right.\right.\right.\nonumber\\ \left.\left.\left.
   +\left(34650 \pi +33600 \pi  \epsilon ^2\right) \xi ^3+129600 \xi
^4+\left(31185 \pi +30240 \pi  \epsilon ^2\right) \xi^5\right]
X\right.\right.\nonumber\\ \left.\left.
   +\left[-12960 \xi -60480 \xi ^3-64800 \xi ^5\right] X^2\right\}+\eta ^2
\left\{-4320 \pi +2240 \pi  \epsilon ^2+5760\xi \right.\right.\nonumber\\
\left.\left.
   +\left(1950 \pi +26000 \pi  \epsilon ^2\right) \xi ^2+60480 \xi
^3+\left(34650 \pi +33600 \pi  \epsilon ^2\right)
\xi^4+\left[5760\right.\right.\right.\nonumber\\ \left.\left.\left.
   +\left(6750 \pi -7920 \pi  \epsilon ^2\right) \xi -23040 \xi ^2+\left(-13500
\pi -37200 \pi  \epsilon ^2\right)\xi ^3-120960 \xi
^4\right.\right.\right.\nonumber\\ \left.\left.\left.
   +\left(-34650 \pi -33600 \pi  \epsilon ^2\right) \xi ^5\right]
X\right.\right.\nonumber\\ \left.\left.
   +\left[-8640 \xi +17280 \xi ^3+60480 \xi^5\right] X^2\right\}
   \right],
\end{eqnarray}
\begin{eqnarray}
\fl b_2=\,\frac{\sqrt{1-\epsilon ^2} }{144 \pi ^2}\eta \left[
\left(-3024 \pi +2703 \pi  \epsilon ^2\right) \xi ^2+3456 \xi ^3+3825 \pi 
\epsilon ^2 \xi ^4+\left[\left(1008 \pi -561 \pi \epsilon ^2\right) \xi
\right.\right.\nonumber\\ \left.\left.
+2304 \xi ^2+\left(3024 \pi -3978 \pi  \epsilon ^2\right) \xi ^3-6912 \xi
^4-3825 \pi  \epsilon ^2 \xi^5\right] X+\left[-1152 \xi\right.\right.\nonumber\\
\left.\left.
 -2304 \xi ^3+3456 \xi ^5\right] X^2+\eta ^2 \left\{1344 \pi -1088 \pi  \epsilon
^2-1536 \xi+\left(5040 \pi \right.\right.\right.\nonumber\\ \left.\left.\left.
 -13430 \pi  \epsilon ^2\right) \xi ^2-16128 \xi ^3-17850 \pi  \epsilon ^2 \xi
^4+\left[-1536+\left(-3024 \pi\right.\right.\right.\right.\nonumber\\
\left.\left.\left.\left.
 +3978 \pi  \epsilon ^2\right) \xi +6144 \xi ^2+\left(-5040 \pi +19380 \pi 
\epsilon ^2\right) \xi ^3+32256 \xi ^4\right.\right.\right.\nonumber\\
\left.\left.\left.
 +17850 \pi \epsilon ^2 \xi ^5\right] X+\left[2304 \xi -4608 \xi ^3-16128 \xi
^5\right] X^2\right\}+\eta ^4 \left\{1088 \pi 
\epsilon^2\right.\right.\nonumber\\ \left.\left.
 +4608 \xi +12495 \pi  \epsilon ^2 \xi ^2+17280 \xi ^3+16065 \pi  \epsilon ^2
\xi ^4+\left[-1536-3825 \pi  \epsilon ^2 \xi\right.\right.\right.\nonumber\\
\left.\left.\left.
 -20736 \xi ^2-17850 \pi  \epsilon ^2 \xi ^3-34560 \xi ^4-16065 \pi  \epsilon ^2
\xi ^5\right] X\right.\right.\nonumber\\ \left.\left.
 +\left[3456 \xi +16128 \xi^3+17280 \xi ^5\right] X^2\right\}
   \right],
\end{eqnarray}
\begin{eqnarray}
\fl A^*_2=\,\frac{\epsilon  \sqrt{1-\epsilon ^2} }{576 \pi ^2}\left(1-\eta
^2\right) \left[
-1536+\left(423 \pi -72 \pi  \epsilon ^2\right) \xi +288 \xi ^2+\left(840 \pi
-632 \pi  \epsilon ^2\right) \xi ^3\right.\nonumber\\ \left.
-864  \xi^4+\left(-495 \pi -480 \pi  \epsilon ^2\right) \xi ^5+\left[423 \pi -72
\pi  \epsilon ^2+576 \xi +\left(-747 \pi\right.\right.\right.\nonumber\\
\left.\left.\left.
 +240 \pi \epsilon ^2\right) \xi ^2+\left(-675 \pi +792 \pi  \epsilon ^2\right)
\xi ^4+1728 \xi ^5+\left(495 \pi +480 \pi  \epsilon^2\right) \xi ^6\right]
X\right.\nonumber\\ \left.
 +\left[288+864 \xi ^2-288 \xi ^4-864 \xi ^6\right] X^2+\eta ^4
\left\{-1536+\left(-3729 \pi\right.\right.\right.\nonumber\\ \left.\left.\left.
  -3616 \pi \epsilon ^2\right) \xi -19296 \xi ^2+\left(-13860 \pi -13440 \pi 
\epsilon ^2\right) \xi ^3-21600 \xi ^4\right.\right.\nonumber\\ \left.\left.
  +\left(-10395 \pi-10080 \pi  \epsilon ^2\right) \xi ^5+\left[495 \pi +480 \pi 
\epsilon ^2+12096 \xi +\left(7425 \pi \right.\right.\right.\right.\nonumber\\
\left.\left.\left.\left. 
  +7200 \pi  \epsilon^2\right) \xi ^2+52992 \xi ^3+\left(17325 \pi +16800 \pi 
\epsilon ^2\right) \xi ^4+43200 \xi ^5+\left(10395
\pi\right.\right.\right.\right.\nonumber\\ \left.\left.\left.\left.
   +10080 \pi \epsilon ^2\right) \xi ^6\right] X+\left[-864-12960 \xi ^2-33696
\xi ^4-21600 \xi ^6\right] X^2\right\}\right.\nonumber\\ \left.
   +\eta ^2\left\{-1536+\left(-4014 \pi +2376 \pi  \epsilon ^2\right) \xi +4032
\xi ^2+\left(1740 \pi +8920 \pi  \epsilon ^2\right)
\xi^3\right.\right.\nonumber\\ \left.\left.
   +12096 \xi ^4+\left(6930 \pi +6720 \pi  \epsilon ^2\right) \xi ^5+\left[1170
\pi -312 \pi  \epsilon ^2+3456 \xi +\left(4050\pi
\right.\right.\right.\right.\nonumber\\ \left.\left.\left.\left.
   -4752 \pi  \epsilon ^2\right) \xi ^2-16128 \xi ^3+\left(-4050 \pi -11160 \pi 
\epsilon ^2\right) \xi ^4-24192 \xi^5 \right.\right.\right.\nonumber\\
\left.\left.\left.
   +\left(-6930 \pi -6720 \pi  \epsilon ^2\right) \xi ^6\right]
X\right.\right.\nonumber\\ \left.\left.
   +\left[-576-576 \xi ^2+12096 \xi ^4+12096 \xi ^6\right]X^2\right\}
   \right],
\end{eqnarray}
\begin{eqnarray}
\fl a^*_2=\,\frac{\sqrt{1-\epsilon ^2} }{288 \pi ^2}\left(1-\eta ^2\right)\left[
-2048+512 \epsilon ^2+\left(504 \pi -153 \pi  \epsilon ^2\right) \xi
+\left(1152-288 \epsilon ^2\right) \xi ^2\right.\nonumber\\ \left.
+\left(1512 \pi-1564 \pi  \epsilon ^2\right) \xi ^3+\left(1152-288 \epsilon
^2\right) \xi ^4-1275 \pi  \epsilon ^2 \xi ^5+\left[504
\pi\right.\right.\nonumber\\ \left.\left.
 -153\pi  \epsilon ^2+\left(2304-576 \epsilon ^2\right) \xi +\left(-1008 \pi
+561 \pi  \epsilon ^2\right) \xi ^2+\left(-1512 \pi
\right.\right.\right.\nonumber\\ \left.\left.\left.
 +1989 \pi  \epsilon ^2\right) \xi ^4+\left(-2304+576 \epsilon ^2\right) \xi
^5+1275 \pi  \epsilon ^2 \xi ^6\right]X+\left[1152-288 \epsilon
^2\right.\right.\nonumber\\ \left.\left.
 +\left(-1152+288 \epsilon ^2\right) \xi ^2+\left(-1152+288 \epsilon^2\right)
\xi^4+\left(1152-288 \epsilon ^2\right)\xi ^6\right] X^2\right.\nonumber\\
\left.
 +\eta ^4 \left\{-2048+512 \epsilon ^2-9605 \pi   \epsilon ^2
\xi+\left(4224-1056 \epsilon ^2\right) \xi ^2-35700 \pi  \epsilon ^2 \xi
^3\right.\right.\nonumber\\ \left.\left.
 +\left(5760-1440 \epsilon  ^2\right) \xi ^4-26775 \pi \epsilon ^2 \xi
^5+\left[1275 \pi  \epsilon ^2+\left(-768+192 \epsilon ^2\right)
\xi\right.\right.\right.\nonumber\\ \left.\left.\left.
  +19125 \pi  \epsilon ^2 \xi^2+\left(-12288+3072 \epsilon ^2\right) \xi
^3+44625 \pi  \epsilon ^2 \xi ^4+\left(-11520
\right.\right.\right.\right.\nonumber\\ \left.\left.\left.\left.
  +2880  \epsilon ^2\right) \xi ^5+26775\pi  \epsilon ^2 \xi ^6\right]
X+\left[1152-288 \epsilon ^2+\left(3456-864 \epsilon ^2\right) \xi
^2\right.\right.\right.\nonumber\\ \left.\left.\left.
  +\left(8064-2016 \epsilon^2\right) \xi ^4+\left(5760-1440 \epsilon ^2\right)
\xi ^6\right] X^2\right\}+\eta ^2 \left\{-2048+512 \epsilon
^2\right.\right.\nonumber\\ \left.\left.
  +\left(-6552\pi +5814 \pi  \epsilon ^2\right) \xi +\left(-6912+1728 \epsilon
^2\right) \xi ^2+\left(-7560 \pi \right.\right.\right.\nonumber\\
\left.\left.\left.
   +23120 \pi  \epsilon^2\right) \xi ^3+\left(-2304+576 \epsilon ^2\right) \xi
^4+17850 \pi  \epsilon ^2 \xi ^5+\left[1512 \pi -714 \pi 
\epsilon^2\right.\right.\right.\nonumber\\ \left.\left.\left.
   +\left(-4608+1152 \epsilon ^2\right) \xi +\left(9072 \pi -11934 \pi  \epsilon
^2\right) \xi ^2+\left(7560 \pi -29070 \pi \epsilon ^2\right) \xi ^4
\right.\right.\right.\nonumber\\ \left.\left.\left.
   +\left(4608-1152 \epsilon ^2\right) \xi ^5-17850 \pi  \epsilon ^2 \xi
^6\right] X+\left[2304-576\epsilon ^2+\left(11520
\right.\right.\right.\right.\nonumber\\ \left.\left.\left.\left.
   -2880 \epsilon ^2\right) \xi ^2+\left(6912-1728 \epsilon ^2\right) \xi
^4+\left(-2304+576 \epsilon^2\right) \xi ^6\right) X^2\right)
   \right].
\end{eqnarray}

\section{Relative binding energy up to the seventh order}

\begin{eqnarray}
\fl E_B^{\rm{(rel)}}=\left(1-\epsilon ^2\right)\left\{\frac{1}{5} \gamma
+\frac{2}{175} \left(7-2 \epsilon ^2\right) \gamma^2+\frac{1}{3024000 \pi
^2}\left[-179200 \left(64-52 \epsilon ^2+3 \epsilon
^4\right)\right.\right.\nonumber\\ \left.\left.
+\pi ^2\left(1275648-1042821 \epsilon ^2+58618 \epsilon ^4\right)\right] \gamma
^3\right.\nonumber\\ \left.
-\frac{1}{1192181760000 \pi ^2} \left[-1433600 \left(-4100096+4180767 \epsilon
^2-513631 \epsilon ^4\right.\right.\right.\nonumber\\ \left.\left.\left.
+63360 \epsilon ^6\right)+\pi ^2\left(-619450073088+643913271759 \epsilon
^2-89452005694 \epsilon ^4\right.\right.\right.\nonumber\\ \left.\left.\left.
+8224514048 \epsilon ^6\right)\right] \gamma^4\right.\nonumber\\ \left.
+\frac{1}{3808877661388800000 \pi^4}\left[601896716861440000 \left(3 \epsilon
^8-79 \epsilon ^6+556 \epsilon ^4 \right.\right.\right.\nonumber\\
\left.\left.\left.
-992 \epsilon ^2+512\right)-1433600 \pi ^2 \left(296565800960
\epsilon^8-5672224097024 \epsilon ^6 \right.\right.\right.\nonumber\\
\left.\left.\left.
+23997689785665 \epsilon ^4-40280188824129 \epsilon
^2+22326414409728\right)\right.\right.\nonumber\\ \left.\left.
+\pi ^4 \left(22758910736859136 \epsilon^8-352585265954279424 \epsilon ^6
\right.\right.\right.\nonumber\\ \left.\left.\left.
+126049642449134226 \epsilon ^4+181545217870667751 \epsilon ^2
\right.\right.\right.\nonumber\\ \left.\left.\left.
+124003773931585536\right)\right]\gamma^5\right.\nonumber\\ \left.
+\frac{1}{157259722042569129984000000 \pi ^4}\left[\pi ^4
\left(-497795548816521583132672 \epsilon ^{10} \right.\right.\right.\nonumber\\
\left.\left.\left.
+8577976113717247992135680 \epsilon^8-79165709064202270230097920 \epsilon^6
\right.\right.\right.\nonumber\\ \left.\left.\left.
+33090553045688299692921150 \epsilon ^4+151318800941422343720072625 \epsilon ^2
\right.\right.\right.\nonumber\\ \left.\left.\left.
-110706740848531490092351488\right)-3064201467658240000\left(17842176 \epsilon
^{10} \right.\right.\right.\nonumber\\ \left.\left.\left.
 -386936448 \epsilon ^8+3563361915 \epsilon ^6-14430944246 \epsilon
^4+20418008059 \epsilon ^2 \right.\right.\right.\nonumber\\ \left.\left.\left.
 -9181331456\right)+1433600\pi ^2 \left(7439589986166374400 \epsilon ^{10}
\right.\right.\right.\nonumber\\ \left.\left.\left.
-141373329772495503360 \epsilon ^8+1305344205610400847104 \epsilon ^6
\right.\right.\right.\nonumber\\ \left.\left.\left.
-3328474247978028922343 \epsilon^4+3357975255903479587047 \epsilon ^2
\right.\right.\right.\nonumber\\ \left.\left.\left.
-1218264268983760846848\right)\right]\gamma^6\right.\nonumber\\ \left.
+\frac{1}{42049186426629786850736209920000000 \pi
^6}\left[\right.\right.\nonumber\\ \left.\left.
-23625990069063869264830660608000000\left(3 \epsilon ^4-52 \epsilon ^2+64\right)
\left(\epsilon ^4-9 \epsilon ^2+8\right)^2 \right.\right.\nonumber\\
\left.\left.
+192414534860800 \pi ^2 \left(138824602612452556800 \epsilon  ^{12}
\right.\right.\right.\nonumber\\ \left.\left.\left.
-3668983313237803008000 \epsilon ^{10}+31662317283432188592128 \epsilon ^8
\right.\right.\right.\nonumber\\ \left.\left.\left.
-128804062549475170301551 \epsilon ^6+254210776514571609476318 \epsilon^4
\right.\right.\right.\nonumber\\ \left.\left.\left.
-232717498273294565021295 \epsilon ^2+79178625735391287705600\right)
\right.\right.\nonumber\\ \left.\left.
-204800 \pi ^4 \left(14969569145984059227242496000 \epsilon^{12}
\right.\right.\right.\nonumber\\ \left.\left.\left.
-309560598463979551791045935104 \epsilon^{10}\right.\right.\right.\nonumber\\
\left.\left.\left.
+1708384495262927033407154421760 \epsilon ^8\right.\right.\right.\nonumber\\
\left.\left.\left.
-6610208901059681863920531779328 \epsilon ^6\right.\right.\right.\nonumber\\
\left.\left.\left.
+10529097859427840958591426694917 \epsilon ^4\right.\right.\right.\nonumber\\
\left.\left.\left.
-6453024674363267467191440465925 \epsilon^2\right.\right.\right.\nonumber\\
\left.\left.\left. +1142467883386445362059993415680\right)
\right.\right.\nonumber\\ \left.\left.+3 \pi ^6
\left(36492994140024750194896213639168 \epsilon
^{12}\right.\right.\right.\nonumber\\ \left.\left.\left.
-597916823876067055030014970953728 \epsilon ^{10}
\right.\right.\right.\nonumber\\ \left.\left.\left.
+1607091649292843235240088719851520 \epsilon  ^8
\right.\right.\right.\nonumber\\ \left.\left.\left.
-15422004663792240785181372461404160 \epsilon ^6
\right.\right.\right.\nonumber\\ \left.\left.\left.
+19532343512100245042235455926381170 \epsilon ^4
\right.\right.\right.\nonumber\\ \left.\left.\left.
+5613107997015131353134697962936567 \epsilon^2 \right.\right.\right.\nonumber\\
\left.\left.\left.
-10612032886678676548237530771750912\right)\right]\gamma^7
\right\} +\mathcal{O}\left(\gamma^8\right).
\end{eqnarray}

\end{document}